\documentclass[prd,twocolumn,superscriptaddress,floatfix,amsmath,amssymb,amsfonts,longbibliography]{revtex4}

\usepackage{float} 

\usepackage{comment}
\usepackage[normalem]{ulem}
\usepackage[english]{babel}
\usepackage{graphicx}
\usepackage{dcolumn}
\usepackage{bm}
\usepackage{blindtext}
\usepackage{slashed}
\usepackage{verbatim}
\usepackage{mathrsfs}
\usepackage{amsmath}
\usepackage{cancel}
\usepackage{physics}
\usepackage{epstopdf}
\usepackage{mathtools}
\usepackage{color}

\newcommand{\ii}{\mathrm{i}}

\usepackage[usenames,dvipsnames]{pstricks}
\usepackage{epsfig}
\usepackage{pst-grad} 
\usepackage{pst-plot} 

\usepackage{hyperref}

\usepackage{verbatim}

\begin{document}

\title{Neutrino flavor oscillations without flavor states}

\author{Bruno de S. L. Torres}
\email{bdesouzaleaotorres@perimeterinstitute.ca}
\affiliation{Perimeter Institute for Theoretical Physics, Waterloo, Ontario, N2L 2Y5, Canada}

\affiliation{Department of Physics and Astronomy, University of Waterloo, Waterloo, ON N2L 3G1, Canada}

\affiliation{Instituto de F\'{i}sica Te\'{o}rica, Universidade Estadual Paulista, S\~{a}o Paulo, S\~{a}o Paulo, 01140-070, Brazil}

\author{T. Rick Perche}
\email{trickperche@perimeterinstitute.ca}
\affiliation{Perimeter Institute for Theoretical Physics, Waterloo, Ontario, N2L 2Y5, Canada}
\affiliation{Department of Physics and Astronomy, University of Waterloo, Waterloo, ON N2L 3G1, Canada}
\affiliation{Instituto de F\'{i}sica Te\'{o}rica, Universidade Estadual Paulista, S\~{a}o Paulo, S\~{a}o Paulo, 01140-070, Brazil}

\author{Andr\'e G.\ S.\ Landulfo}\email{andre.landulfo@ufabc.edu.br}
\affiliation{Centro de Ci\^encias Naturais e Humanas,
Universidade Federal do ABC, Avenida dos Estados, 5001, 09210-580, Santo
Andr\'e, S\~ao Paulo, Brazil}

\author{George E. A. Matsas}
\email{george.matsas@unesp.br}
\affiliation{Instituto de F\'{i}sica Te\'{o}rica, Universidade Estadual Paulista, S\~{a}o Paulo, S\~{a}o Paulo, 01140-070, Brazil}

\begin{abstract}
\textcolor{black}{We analyze the problem of neutrino oscillations via a fermionic particle detector model inspired by the physics of the Fermi theory of weak interactions. The model naturally leads to a description of emission and absorption of neutrinos in terms of localized two-level systems. By explicitly including source and detector as part of the dynamics, the formalism is shown to recover the standard results for neutrino oscillations without mention to ``flavor states'', which are ill-defined in Quantum Field Theory~(QFT). This illustrates how particle detector models  provide a powerful theoretical tool to approach the measurement issue in QFT and emphasizes that the notion of flavor states, although sometimes useful, must not play any crucial role in neutrino phenomenology.}

\end{abstract}

\maketitle

\section{Introduction}

    \textcolor{black}{Neutrinos have become one of the greatest protagonists in the search for hints of physics beyond the Standard Model. It is believed that a better understanding of neutrino physics could shed light into broad, long-standing questions in fundamental physics, which include the nature of dark matter~\cite{BOYARSKY20191} and the asymmetry between matter and antimatter in the universe~\cite{RevModPhys.84.515}. One of the most direct indications that neutrinos provide for the need of extensions of the Standard Model comes from the phenomenon of flavor oscillations, which implies that the neutrinos (that are predicted to be massless in the Standard Model) are actually massive, and that the neutrinos with well-defined flavor---which couple directly to the charged leptons through the weak interactions---are linear combinations of the neutrinos with well-defined mass.}

Despite its apparent simplicity, the description of flavor states as linear combinations of mass ones has raised important questions regarding whether this is well defined within the framework of QFT. Whereas the construction of a Fock basis of massive neutrinos is straightforward via standard canonical quantization, the difficulty in formulating Fock states for neutrinos with well-defined flavor~\cite{Eur.Phys.J.C39} makes it relevant to investigate whether a Fock basis for flavor neutrinos is necessary at all. Indeed, attempts to construct a Fock space for flavor neutrinos~\cite{BLASONE1995283} exhibit undesirable features~\cite{PhysRevD.63.125015} as, {\em e.g.}, the fact that the canonical vacuum state would be populated with flavor neutrinos \textcolor{black}{or that the flavor vacuum state would not be invariant under time translations}. It has been shown, indeed, that flavor states can be defined only under certain conditions depending, in general, on the underlying phenomenological process~\cite{PhysRevD.45.2414}. As a result, the usual description of flavor states as linear combinations of mass neutrino states turns out to be circumstantial rather than a fundamental feature of neutrinos physics (for a more comprehensive discussion see Ref.~\cite{Giunti:2007ry}). In this vein, it would be fruitful to devise a framework where all neutrino phenomenology is entirely rephrased in terms of neutrinos with well-defined mass~\cite{BEUTHE2003105}. In particular, here, we focus on the neutrino oscillation phenomenon.

In order to accomplish this goal, it turns out to be useful to think more thoroughly about the process of emission and detection of neutrinos as an inherent part of the dynamics. This is efficiently achieved in the framework of \emph{particle detector models}. Broadly speaking, particle detectors consist of controllable quantum systems that couple to quantum fields in a localized region of spacetime. It has been shown that particle detectors are a powerful tool in various areas of theoretical physics, which range from quantum optics to QFT in curved spacetimes. They provide an appealing operational approach for the problem of measurement in QFT and have also shed light into a wide array of phenomena, including, but not restricted to, the Unruh and Hawking effects~\cite{PhysRevD.14.870}, entanglement harvesting~\cite{Pozas2016}, and quantum energy teleportation~\cite{teleportation}.

A prototype particle detector model is the Unruh-DeWitt~(UDW) one. This is a localized two-level quantum monopole, which couples linearly to scalar fields. There has been increasing interest, however, in extending this model to consider the coupling with higher-spin fields. For instance, the coupling of a detector to the electromagnetic field has been shown to model interactions of atoms with light. There have also been proposals of \textcolor{black}{particle detector models} coupling to the linearized gravitational field~\cite{remi} with the intent of probing underlying quantum gravitational effects. It is then just natural to consider a \textcolor{black}{detector-based} framework that describes the coupling to a fermionic field. \textcolor{black}{As we will see, it is possible to probe the phenomenon of neutrino flavor oscillations using one such detector.}

In summary, the major purpose of the present paper is to phrase the phenomenology of neutrino oscillations with the explicit use of particle detector models, in a way that naturally precludes the notion of neutrino flavor states. The conceptualization of neutrino oscillations in quantum field theory without flavor states has been previously studied~\cite{Giunti93, Grimus, Grimus2, Naumov_2010, beuthe, Zoltan, Cardall}, but the role of particle detector models in those descriptions was not explored or emphasized. Given the growing importance of particle detector models in theoretical physics, formulating neutrino oscillations with the aid of a fermionic detector model represents relevant progress in our understanding of the phenomenon. It is important to note that Kobach, Manohar, and McGreevy have recently used a preliminary detector-based strategy to analyze the oscillation phenomenon for scalar fields~\cite{Kobach}. Our present approach, however, does not make use of the rotating wave approximation (which may yield important differences, {\em e.g.}, in cases where detector and source couple to the field for finite times~\cite{PhysRevD.48.3731}). Our description based on an explicit interaction action~\eqref{interactionscalar1} seems also better suited to include effects due to relative source-detector motions. Moreover, here, we take a step further and introduce a fermion detector which is necessary to treat the oscillation phenomenon in more realistic terms.

The paper is organized as follows. In Sec.~\ref{secNaive}, we review the usual quantum-mechanical derivation of the oscillation phenomenon based on flavor and mass states. In Sec.~\ref{scalarneutrinos}, we consider the simplified case where fermionic fields with flavor mixing are replaced by scalar ones and show how a suitable UDW model can describe emission and absorption processes of ``scalar'' neutrinos. In particular, it clarifies how the standard picture of flavor oscillations can be rephrased in terms of detector observables. As a result, we obtain an exact quantum-field-theoretical result at the lowest order, which is in agreement with the standard oscillation probability result in the proper regime. In Sec.~\ref{fermionicNeutrinos}, we present a fermionic particle detector model derived in the context of the Fermi theory. We apply our model to the process of emission and absorption of neutrinos, and show under what circumstances it can recover the scalar result. In particular, it is shown that one can successfully account for the phenomenon of flavor oscillations without the need of any notion of flavor states. In Sec.~\ref{conclusion}, we present our final conclusions. 

We will assume metric signature $(+,-,-,-)$ and natural units, $\hbar=c=1$, unless stated otherwise.

\section{Quantum Mechanics approach to Neutrino Oscillations}
\label{secNaive}

In the proper regime, the oscillation phenomenon can be derived using plain quantum mechanics overlooking the neutrino fermionic nature~\cite{1978PhR....41..225B}. We review it here briefly for further comparison. 

Let us denote the state of neutrinos with masses, $m_i,\; i =1,2,3,$ and  momentum $\bm p$ by $\ket{\nu_i(\bm p)}$. These states are regarded as the eigenstates of the free Hamiltonian, $\hat{H}$, so that they satisfy
\begin{equation}
    \hat{H}\ket{\nu_i(\bm p)} =  \omega_i(\bm p) \ket{\nu_i(\bm p)},
\end{equation}
where ${\omega_i(\bm p)} = \sqrt{m_i^2+\bm p^2}$.

Next, let us define the corresponding flavor states, $\ket{\nu_\alpha(\bm p)},\; \alpha =1,2,3$, in terms of the mass states through the Pontecorvo–Maki–Nakagawa–Sakata (PMNS) matrix $U_{\alpha j}$:
\begin{equation}\label{mixingQM}
    \ket{\nu_\alpha(\bm p)} = \sum_j U_{\alpha j} \ket{\nu_j(\bm p)},
\end{equation}
where
\begin{equation}
    \sum_j U_{\alpha j} U_{\beta j}^* = \delta_{\alpha\, \beta},  \quad
    \sum_\alpha U_{\alpha j} U_{\alpha k}^* = \delta_{j\, k}.
\end{equation}
Flavor states will be labeled by Greek indices $\alpha,\beta = e,\mu,\tau$, corresponding to the  electron, muon, and tau neutrinos. 

In this context, the neutrino oscillation phenomenon is associated with the nonconservation of the neutrino flavor between production and detection. Indeed, given that the massive neutrinos are eigenstates of the Hamiltonian, each one of them evolves in time with a global phase: $\ket{\nu_i(t,\bm p)} = e^{-\ii \omega_i(\bm p)t}\ket{\nu_i(\bm p)}$. It follows, then, that the time evolution of the flavor neutrinos is given by
\begin{equation}
    \ket{\nu_\alpha(t,\bm p)} 
    = \sum_j U_{\alpha j} e^{-\ii \omega_j(\bm p) t}\ket{\nu_j(\bm p)}.
\end{equation}
The amplitude associated with emitting $|\nu_\alpha (\boldsymbol{p}) \rangle$ and measuring $|\nu_\beta (\boldsymbol{p}) \rangle$ at some later time $t$ is
\begin{eqnarray}\label{amplitudeProvisoria}
    \mathcal{A}_{\alpha\rightarrow \beta}(t) 
    &=& 
    \langle \nu_\beta(\bm p) | \nu_\alpha(t,\bm p) \rangle
    \nonumber \\
    &=& 
    \sum_{i j} U_{\alpha j}U_{\beta i}^{\ast}  e^{-\ii\omega_j(\bm p) t}\braket{\nu_i(\bm p)}{\nu_j(\bm p)}
    \nonumber \\
    &=& 
    \sum_{j} U_{\alpha j}U_{\beta j}^{\ast}  e^{-\ii \omega_j(\bm p) t}.
\end{eqnarray}
Given that neutrinos are ultrarelativistic, $m_i \ll |\boldsymbol{p}| \equiv p$, in oscillation experiments, the phases acquired by the different mass neutrinos can be cast as
\begin{equation}
    \omega_j(\bm p)t 
    \approx 
    pL+ \frac{m_j^2}{2p}L,
\end{equation}
where $L\approx t$ is the distance travelled by the neutrino in the time interval $t$. Hence, Eq.~\eqref{amplitudeProvisoria} becomes
\begin{equation}
    \mathcal{A}_{\alpha\rightarrow \beta} (L) 
    \approx e^{-\ii pL}  
    \sum_{j} U_{\alpha j}U_{\beta j}^{\ast}  e^{-\ii m_j^2L/2p},
\end{equation}
which yields the following probability for the process
\begin{equation}\label{basic}
    \mathcal{P}_{\alpha\rightarrow\beta} 
    \approx 
    \abs{\sum_{j} U_{\alpha j}U_{\beta j}^{\ast}  e^{-\ii m_j^2L/2p}}^2.
\end{equation}
Equation~(\ref{basic}) agrees with every neutrino oscillation experiment to date. This encodes the essence of the phenomenon, where the phase difference acquired between different mass neutrinos is proportional to the difference between the squared masses. 

It is remarkable how much can be obtained from standard quantum mechanics neglecting the more detailed properties of the particles under consideration. A more fundamental description of the phenomenon, however, begs for a QFT analysis. This is necessary, {\em e.g}, to make a precise sense of the ``$\, \approx \, $'' symbols introduced in the derivation above.

In the QFT context, the mass neutrinos are associated with quantum fields $\hat{\nu}_i(x)$, for which one can define creation and annihilation operators associated with massive neutrinos and antineutrinos of masses $m_i$ through canonical quantization. The states  $\ket{\nu_i(\bm p)}$  should be identified with the one-particle states in the Fock space for their respective fields. The mixing is now encoded in the fields rather than in the states:
 \begin{equation}\label{mixingQFT}
     \hat{\nu}_\alpha(x) \equiv \sum_j U_{\alpha j}\hat{\nu}_j(x),
 \end{equation}
 where $\hat{\nu}_\alpha({x})$ is seen as the quantum field associated with the $\alpha$-flavor neutrino. However, the spectrum of real particles of the theory is only well understood with respect to the fields with well-defined mass. The states $\ket{\nu_\alpha(\bm p)}$ defined in Eq.~\eqref{mixingQM} do not correspond to Fock states of the  flavor neutrino fields, and should instead be interpreted as convenient phenomenological states~\cite{Giunti:2007ry}. Therefore, the quantum-mechanical approach described in this section should be seen, at best, as a good first approximation.
 
 The next sections will be devoted to showing how the neutrino oscillation phenomenon can be fully understood within the framework of QFT endowed with a suitable neutrino detector, which will allow us to include the neutrino production and detection in the analysis.


\section{Scalar Neutrino Oscillations via Unruh-DeWitt Detectors}
\label{scalarneutrinos}

In a very broad sense, the conceptualization of a theoretical framework for neutrino oscillations that does not evoke flavor states can naturally be accommodated by a more careful analysis of how flavor oscillations are measured. One way of obtaining such a framework is by employing particle detectors, which are very well adapted to the intuitive picture of localized emission and absorption of particles. Moreover, from a fundamental point of view, measurements in QFT are better understood in terms of particle detectors. The most famous detector model is the UDW one. It consists of a first-quantized two-level system whose internal structure can, to a good approximation, be described by a nonrelativistic system, which couples linearly to a real scalar field in a localized region of space and time.


The interaction action of a pointlike UDW detector with monopole moment $\hat{\mu}(\tau)$ coupled to a real scalar field $\hat{\phi}({x})$ is
\begin{equation}\label{SIE}
    S_I = -\lambda \int_{-\infty}^{+\infty}
    \dd\tau \chi(\tau) \hat{\mu}(\tau) \hat{\phi}(\mathsf{x} (\tau)).
\end{equation}
Here, $\lambda$ is a coupling constant, $\mathsf{x} (\tau)$ denotes the trajectory of the detector parametrized by its proper time $\tau$, and $\chi(\tau)$ is a \emph{switching function} that is responsible for dictating the temporal profile of the interaction strength. Let us define (in the interaction picture)
\begin{equation}\label{mu(tau)}
    \hat{\mu}(\tau) \equiv \hat \sigma^+ (\tau) + \sigma^- (\tau)
\end{equation}
with  
$\hat \sigma^\pm (\tau) \equiv e^{\pm \ii \Omega \tau}\hat \sigma^\pm$,
where
\begin{equation}\label{sigma+-(tau)}
    \hat{\sigma}^+ = \ket{e}\!\!\bra{g}\quad {\rm and} \quad
    \hat{\sigma}^- = \ket{g}\!\!\bra{e},
\end{equation}
are raising and lowering operators, respectively, connecting the  ground, $\ket{g}$, and excited, $\ket{e}$, states of the two-level system and $\Omega$ is the corresponding proper energy gap. For later comparison, let us recast Eq.~(\ref{SIE}) as
\begin{equation}\label{SIE2}
    S_I = -\lambda \int_{-\infty}^{+\infty}
    \dd\tau \, \chi(\tau) \, [\hat{\sigma}^+ (\tau) \hat{\phi} (\mathsf{x} (\tau))+\hat{\sigma}^- (\tau) \hat{\phi} (\mathsf{x} (\tau))].
\end{equation}

Now, let us write the interaction action for a (scalar) neutrino coupled with a source, $s$, and detector, $d$, both of them modeled by UDW two-level systems as
\begin{eqnarray}\label{interactionscalar1}
    S_I 
    =
    &-&\lambda_s\int_{-\infty}^{+\infty}  
    \dd \tau_s\chi_s(\tau_s)\hat{\mu}_s(\tau_s)\hat{\phi}_\alpha(\mathsf{x}_s(\tau_s)) 
    \nonumber \\
    &-& \lambda_d\int_{-\infty}^{+\infty} 
    \dd \tau_d\chi_d(\tau_d)\hat{\mu}_d(\tau_d)\hat{\phi}_\beta(\mathsf{x}_d(\tau_d)),
 \end{eqnarray}  
where $\tau_s$ and $\tau_d$ are the source and detector proper times, respectively. We are considering a setup where the source and detector couple to  flavor neutrino fields $\hat{\phi}_\alpha(x)$ and $\hat{\phi}_\beta(x)$, respectively.
They are defined as linear combinations of the neutrino fields with well-defined mass, $\hat{\phi}_j(x)$, as
\begin{equation}
    \hat\phi_\zeta(x) = \sum_j U_{\zeta j} \hat\phi_j(x ),\quad
   \zeta=\alpha, \beta,
\end{equation}
where the PMNS matrix $U_{\zeta j}$ is real in the scalar case. For the sake of further comparison, let us use  Eq.~(\ref{mu(tau)}) to rewrite Eq.~(\ref{interactionscalar1}) as
\begin{eqnarray} \label{interactionscalar2}
    S_I
    =
    &-&\lambda_s\int_{-\infty}^{+\infty} \!\!\!\!
    \dd \tau_s\chi_s(\tau_s)\hat{\sigma}^-_s(\tau_s)\hat{\phi}_\alpha(\mathsf{x}_s(\tau_s)) \! + \!\textrm{H.c.}
    \nonumber \\
    &-& \lambda_d\int_{-\infty}^{+\infty} \!\!\!\!
    \dd \tau_d\chi_d(\tau_d)\hat{\sigma}^-_d(\tau_d)\hat{\phi}_\beta(\mathsf{x}_d(\tau_d))\! + \!\textrm{H.c.}
\end{eqnarray}

The process which will represent an oscillation experiment has, thus, the following initial and final states
\begin{equation}
    \ket{i} = \ket{0}\ket{e_s}\ket{g_d} \quad 
    {\rm and}\quad 
    \ket{f} = \ket{0}\ket{g_s}\ket{e_d},
\end{equation}
respectively, where $\ket{0}$ is the vacuum state of the three neutrino mass fields ({\em i.e.}, the state that is annihilated by all the annihilation operators of the mass neutrino fields). 

\begin{figure}
\begin{center}
\includegraphics[width=8 truecm]{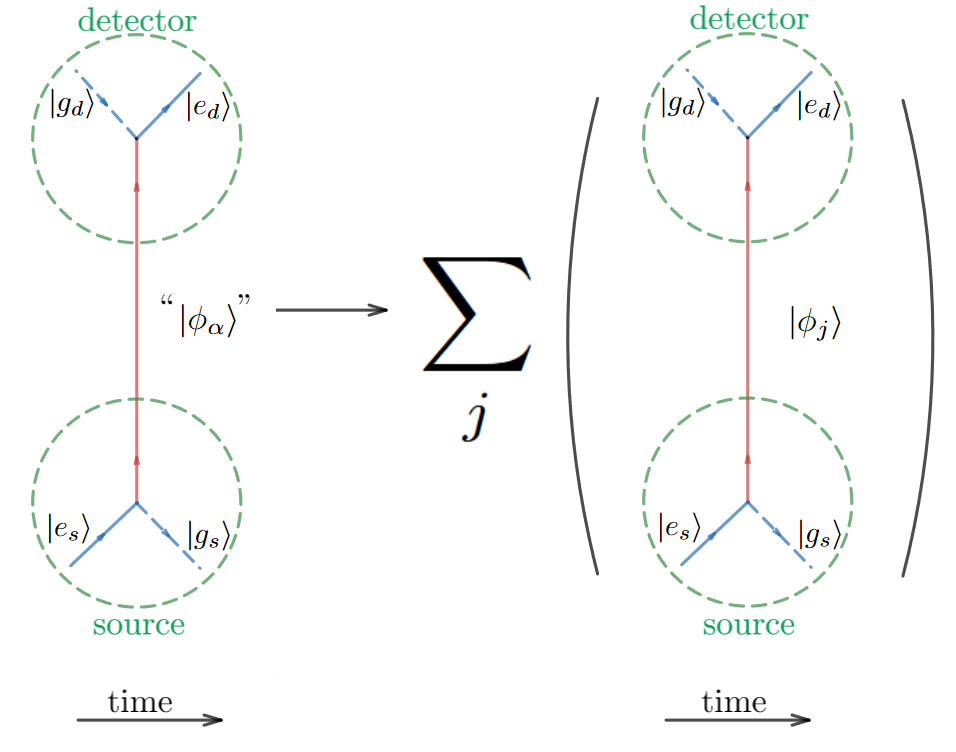}
\end{center}
\caption{The figure illustrates our understanding of the oscillation process. In the naive quantum mechanical approach a flavor neutrino ``$|\phi_\alpha \rangle$" is emitted due to some source deexcitation and oscillates up to the detection moment. (We are assuming that the source-detector distance is not larger than the decoherence length.) In the present approach, the source deexcitation emits coherently three (well-defined) mass neutrinos which eventually excite the detector. The design of the oscillation experiment is codified in the $\alpha$ and $\beta$ flavor labels present in the interaction action~(\ref{interactionscalar2}). For the DUNE, {\em e.g.}, $\alpha = \mu$ and $\beta= e, \mu, \tau$.} 
\label{FeynmanESC}
\end{figure}

Thus, the oscillation event is rephrased in terms of states associated with the source and detector with no intermediate state for the neutrino fields themselves being assumed -- see Fig.~\ref{FeynmanESC}. What would be otherwise interpreted as the {\em ``emission of an $\alpha$-neutrino''} and {\em ``detection of a $\beta$-neutrino''} is understood now as the {\em ``deexcitation of an $\alpha$-source"} and {\em 
``excitation of a $\beta$-detector"}, respectively. 

Assuming that both source and detector follow inertial trajectories at rest with respect to each other, there is a Cartesian coordinate system $(t,\boldsymbol{x})$ where the worldlines can be parametrized by
\begin{equation}\label{inertialtrajectories}
    \mathsf{x}_s(\tau_s) = (t, \boldsymbol{0})
    \quad {\rm and} \quad 
    \mathsf{x}_d(\tau_d) = (t, \boldsymbol{L}),
\end{equation}
with $|\boldsymbol{L}|\equiv L= {\rm const}$ being the distance between source and detector. 

It is straightforward to read the corresponding interaction Hamiltonian from Eq.~(\ref{interactionscalar2}):
\begin{align}
\hat{V}_I(t) 
    & = 
    \underbrace{
    \lambda_s\chi_s(t)\hat{\sigma}^-_s(t)\hat{\phi}_{\alpha}(\mathsf{x}_s) + {\rm H.c.}
    }_\text{$ = \hat{V}_s(\mathsf{x}_s)$
    } 
    \nonumber \\
    & + 
    \underbrace{
    \lambda_d\chi_d(t)\hat{\sigma}^-_d(t)\hat{\phi}_{\beta}(\mathsf{x}_d) + {\rm H.c.}
    }_\text{$=\hat{V}_d(\mathsf{x}_d)$
    }
\end{align}
and so, up to the lowest order in perturbation theory, the oscillation amplitude is 
\begin{eqnarray*}
    \mathcal{A_{\alpha\rightarrow\beta}} 
    &=&  
    \bra{f}\mathcal{T}\exp(\ii S_I)\ket{i}  
    \nonumber \\
    &=&
    - \int_{-\infty}^{+\infty} \!\!\!\! \dd t 
      \int_{-\infty}^{t}\!\!\!\! \dd t' 
    \bra{f} \hat{V}_s(t)\hat{V}_d(t') + \hat{V}_d(t)\hat{V}_s(t') \ket{i}, 
\end{eqnarray*}
where $\mathcal{T}$ is the time-ordering operator. By evaluating it, we obtain    
    \begin{eqnarray}\label{intermediateamplitudescalar}
    \mathcal{A_{\alpha\rightarrow\beta}} 
    &=&
    - \lambda_s\lambda_d \sum_{j}U_{\alpha j}U_{\beta j} 
    \int \dfrac{\dd^3 p_j}{(2\pi)^3} \dfrac{1}{2\omega_j(\boldsymbol{p})}
    \nonumber \\
    &\times&
    \left[ 
    F_j(\bm{p})e^{-\ii \boldsymbol{p}_j\cdot\boldsymbol{L}} +G_j(\bm{p}) e^{\ii \boldsymbol{p}_j\cdot\boldsymbol{L}}
    \right],
\end{eqnarray}
where
\begin{eqnarray}
    F_j(\bm{p}) 
    &\equiv& 
    \int_{-\infty}^{+\infty}\!\!\!\dd t \int_{-\infty}^{t}\!\!\!\dd t' \chi_s(t)\chi_d(t')
    \nonumber \\
    &\times& e^{\ii(\Omega_d + \omega_j(\boldsymbol{p}))t'}e^{-\ii (\Omega_s + \omega_j(\boldsymbol{p}))t}, 
    \label{defF}
\end{eqnarray}  
\begin{eqnarray}
    G_j(\bm{p}) 
    &\equiv& 
    \int_{-\infty}^{+\infty}\!\!\!\dd t \int_{-\infty}^{t}\!\!\!\dd t'\chi_d(t)\chi_s(t')
    \nonumber \\
    &\times& e^{-\ii (\Omega_s - \omega_j(\boldsymbol{p}))t'}e^{\ii (\Omega_d - \omega_j(\boldsymbol{p}))t}, \label{defG}
\end{eqnarray}
$\omega_j(\bm{p}) = \sqrt{|\bm{p}_j|^2 + m_j^2}$, and we have used
\begin{equation}
    \bra{0}\phi_j(x)\phi_k(x')\ket{0} = 
    \delta_{jk}
    \int \frac{\dd^3 p_j}{16 \pi^3 \omega_j(\bm{p})}
    e^{-\ii p_j\cdot (x - x')}.
\end{equation}
(We would rather use $\omega_j (\boldsymbol{p})$ than $\omega_j (\boldsymbol{p}_j)$ to simplify the notation.)

In order to make further progress, we choose a particular temporal profile for the source and detector interactions with the field. Our source will remain coupled to the field for an arbitrarily long time, whereas the detector is kept switched on during the time interval $\Delta t \equiv t_1-t_0 >0$. The following switching functions model quite well this setup:
\begin{equation}\label{switching}
    \chi_s(t) = e^{-\epsilon|t|}, 
    \quad
    \chi_d(t) = \Theta(t - t_0) - \Theta(t - t_1),
\end{equation}
where $\Theta(t)$ is the Heaviside step function, and $\epsilon$ is a small regulator introduced to guaranty that the integrals converge at $t\to \pm\infty$. 

\begin{figure}
\begin{center}
\includegraphics[width=8 truecm]{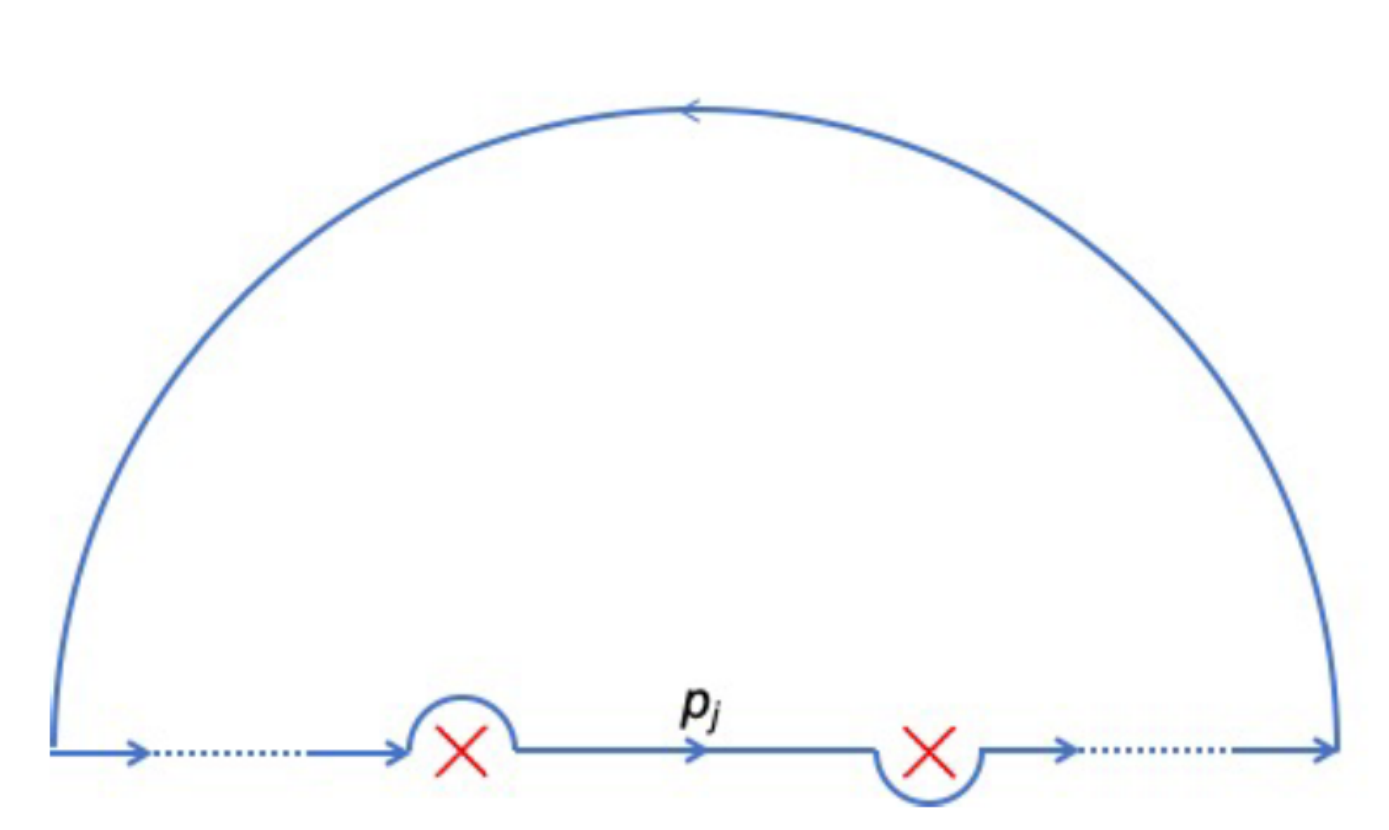}
\end{center}
\caption{Contour of integration used to evaluate Eq.~(\ref{amplitudedaintegral}). The two poles in the real axis are at $ p_j= \mp \Delta_j$ with $\Delta_j \equiv  \sqrt{\Omega_s^2 - m_j^2}$. } 
\label{integration}
\end{figure}

By \textcolor{black}{using} Eq.~(\ref{switching}) to calculate Eqs.~(\ref{defF})-(\ref{defG}), we obtain
\begin{eqnarray}
    F_j(\bm p) &=& -\frac{\Delta T}{\Omega_s+ \omega_j(\bm p)}, \label{resultF}\\
    G_j(\bm p) &=& \frac{\Delta T}{\Omega_s - \omega_j(\bm p)},\label{resultG}
\end{eqnarray}
where 
\begin{eqnarray}
    \Delta T 
    &\equiv& 
    \frac{e^{\ii \Delta\Omega t_1}-e^{\ii \Delta\Omega t_0}}{\Delta\Omega}
    \nonumber \\
    &=&
    2\ii \exp[\ii(\Delta\Omega(t_0 + t_1)/2] \dfrac{\sin{(\Delta\Omega\Delta t/2)}}{\Delta\Omega},
    \label{DeltaT}
\end{eqnarray}
$\Delta\Omega \equiv \Omega_d - \Omega_s$,
and we have taken $\epsilon \rightarrow 0$ at the end. Using Eqs.~(\ref{resultF})--(\ref{resultG}) in Eq.~(\ref{intermediateamplitudescalar}), we write the oscillation amplitude as
\begin{eqnarray}\label{d3pscalar}
    \mathcal{A}_{\alpha\rightarrow\beta} 
    &=& 
    -\lambda_s\lambda_d \Delta T\sum_{j}U_{\alpha j}U_{\beta j} 
    \int\dfrac{\dd^3 p_j}{(2\pi)^3}
    \dfrac{e^{\ii \boldsymbol{p}_j\cdot\boldsymbol{L}}}{\Omega_s^2 - \omega_j^2({\bm p})}
    \nonumber\\
    &=&
    \ii \frac{ \lambda_s\lambda_d}{4\pi^2 L} 
    \Delta T\sum_{j}U_{\alpha j}U_{\beta j}\!\!\! 
    \int_{-\infty}^{+\infty} \!\!\!\!\! 
    \dd p_j\dfrac{ p_j \,e^{\ii p_j L}}{\Omega_s^2 - \omega^2_j(p)}, \label{amplitudedaintegral}
\end{eqnarray}
where $p_j \equiv |\boldsymbol{p}_j|$ and $L \equiv |\boldsymbol{L}_j|$. The integral above can be solved via the residue theorem  once the poles in the real axis are properly circumvented (see Fig.~\ref{integration}). As a result, one obtains
\begin{equation}
    \mathcal{A}_{\alpha\rightarrow\beta} 
    =
    \frac{ \lambda^2}{4\pi L} \Delta T\sum_{j}U_{\alpha j}U_{\beta j}e^{\ii \Delta_j L},
    \label{amplitudeprefinal}
\end{equation}
where $\Delta_j \equiv  \sqrt{\Omega_s^2 - m_j^2}$ and we have defined $\lambda^2 \equiv \lambda_s\lambda_d$.

Thus, using Eq.~(\ref{DeltaT}) in Eq.~(\ref{amplitudeprefinal}), we can write the probability of an ``$\alpha$-flavor'' neutrino to be detected as a ``$\beta$-flavor'' neutrino as 
\begin{equation}\label{probabilidadealfabeta}
        |\mathcal{A}_{\alpha\rightarrow\beta}|^2 
       \! =
       \! \dfrac{\lambda^4}{16\pi^2 L^2 }
       \! \left[\dfrac{\sin{(\Delta\Omega\Delta t/2)}}{\Delta\Omega/2}\right]^2
       \!\left|\sum_{j}U_{\alpha j}U_{\beta j}e^{\ii \Delta_j L}\right|^2 \!\!\!.
\end{equation}
The quotation marks introduced above stress that by $\alpha$-flavor ($\beta$-flavor) neutrino we actually mean a neutrino ``produced (detected) through a weak process involving the corresponding $\alpha$-lepton ($\beta$-lepton).'' Note that the overall prefactor $1/L^2$ in Eq.~\eqref{probabilidadealfabeta} is expected since it \textcolor{black}{encodes} the area-law decay associated with isotropically-emitting sources.

Next, let us consider the particular case where the detector is kept turned on for an arbitrarily long time interval. In this case,
\begin{equation}\label{delta}
    \lim_{\Delta t \rightarrow +\infty}
    \left[\dfrac{\sin{(\Delta\Omega\Delta t/2)}}{\Delta\Omega/2}\right]^2 
    = 
    2\pi \delta(\Omega_d - \Omega_s)\Delta t,
\end{equation}
where $\Delta t$ is associated with the (arbitrary long) detector proper time. Using Eq.~(\ref{delta})  in Eq.~(\ref{probabilidadealfabeta}), we obtain the correspoding (stationary) detector excitation rate as 
\begin{eqnarray}\label{finalprobscalar}
    \Gamma_{\alpha \to \beta} 
    &\equiv& 
    \lim_{\Delta t \rightarrow + \infty} 
    \dfrac{|\mathcal{A}_{\alpha\rightarrow\beta}|^2}{\Delta t}
    \nonumber \\ 
    &=& 
    \dfrac{\lambda^4}{8\pi L^2}\delta(\Omega_d - \Omega_s)\Big|\sum_{j}U_{\alpha j}U_{\beta j}e^{\ii\Delta_j L}\Big|^2.
\end{eqnarray}
The delta function guarantees energy conservation in the ideal case where the detector is kept permanently switched on (no external agent is present making work to turn on/off the detector).

Finally, the familiar Eq.~(\ref{basic}) can be recovered after one normalizes Eq.~(\ref{finalprobscalar}): 
\begin{equation}
\mathcal{P}_{\alpha\rightarrow\beta} 
= 
\frac{\Gamma_{\alpha\to\beta} }{\sum_{\beta' }\Gamma_{\alpha\to\beta'}}
=
\left| 
\sum_{j} U_{\alpha \, j} U_{\beta \, j}  e^{\ii \Delta_j L }
\right|^2.
\label{finalescalar}
\end{equation}
This is an exact result up to the lowest order. In order to recover the approximate result~(\ref{basic}), we must impose that the source produces ultrarelativistic neutrinos: $m_j\ll \Omega_s$, in which  case
\begin{equation}
    \Delta_j  
    \approx 
    \Omega_s  - \dfrac{m_j^2}{2\Omega_s}.
\end{equation}

It should be clear, therefore, that the usual UDW scalar model  is capable of capturing the main features of neutrino oscillations without any assumption about the form of the neutrino fields as they travel from the source to the detector. In particular, we did not have to prescribe by hand any wave packet profile for the emitted neutrinos and, most remarkably, we never had to assume that the states describing the neutrinos participating in the coupling with source and detector were ``flavor eigenstates''. The phenomenon can be understood entirely in terms of Fock states associated with the neutrino fields with well-defined mass. It is also worth pointing out that the energy-momentum uncertainty in emission and detection---necessary for the coherent emission and absorption of massive neutrinos with different mass, and usually described directly in terms of of wave packets---is here realized by the pointlike localization of the coupling in space. This directly leads to infinite uncertainty in the spatial momentum of the neutrino that couples to source and detector. Moreover,  we  could  have  included  temporal profiles with finite spread, where the energy uncertainty would then be directly linked to (the inverse of) the time duration of the interaction. Similarly, a more refined uncertainty in spatial momentum could be explored by  endowing  detector  and  source  with  some  nontrivial spatial  extension,  which  would  then  lead  to  the  use  of \emph{smeared} detector models~\cite{generalrelativistic}.

\section{Neutrino Oscillations via Fermionic Particle Detectors}\label{fermionicNeutrinos}

We will now generalize the scalar UDW model by coupling our two-level system to fermionic fields, which will allow us to properly treat the neutrino as a spinor. Section~\ref{model} below is dedicated to motivating and presenting this new particle detector model. Then, in Sec.~\ref{Fermionic oscillation probability} we use it to investigate the oscillation phenomenon.

\subsection{The model}
\label{model}

The paradigmatic process which will motivate our fermion detector comes from $\beta$-decay: 
    \begin{equation}\label{protonDecay}
        n \rightarrow p + e^- + \bar{\nu}_e.
    \end{equation}
At energies much lower than the $W$- and $Z$-boson masses, this process is well described by the Fermi theory of the weak interactions, where the proton~($p$), neutron~($n$), electron~($e$), and neutrino~($\nu$) fields interact via a pointlike current-current coupling. We will take the Lagrangian that describes the process to be
    \begin{equation}\label{FermiInt}
        \mathcal{L}_{\text{4F}} = -\dfrac{G_F}{\sqrt{2}}\Big[\left(\hat{\bar{\nu}}_e\gamma^\mu (1 - \gamma^5)\hat{e}\right)\left(\hat{\bar{n}}\gamma_\mu (1 - \gamma^5)\hat{p}\right) + \text{H.c.}\Big],
    \end{equation}
where $G_F = 1.17\times 10^{-5}~{\rm GeV}^{-2}$ is the \emph{Fermi constant} and all fields are treated as massive Dirac fermions. The projection operator \mbox{$P_L \equiv ({1}/{2})(1 - \gamma^5)$} appears because charged-current in weak interactions only couple to left-chiral fermions, as is well known. The Lagrangian~\eqref{FermiInt} is motivated by taking the charged-current interaction term in the Fermi theory (which explicitly involves the coupling between electrons, neutrinos, and quarks) and replacing the up and down quarks by the proton and the neutron, respectively. The heuristics behind this is the picture of the transition $n \rightarrow p$ as implicitly $udd \rightarrow uud$. The actual effective Lagrangian that better describes this process is in fact more complicated than this~\cite{betadecay}, but such additional complications will be mostly irrelevant for our following discussion.
    
The first main idea to devise our particle detector is to think of the neutron-proton as quantum states of a localized two-level {\em nucleon} to be described through a semiclassical current. This has been explored elsewhere~\cite{Vanzella2} but here we show how this can be naturally motivated from the 4-fermion theory itself. After this, we will be able to see a transition neutron~$\to$~proton as a nucleon deexcitation, while the emitted neutrino will induce the reverse process, proton~$\to$~neutron, at the detector.

In order to build the raising and lowering operators acting on the nucleon Hilbert space, let us begin recalling the free Dirac field expansion:
    \newcommand{\mf}{\mathsf}
    \begin{eqnarray}
        \hat{f} (x) 
        &=& 
        \sum_{s=\pm 1}\int \frac{\dd^3\bm k}{\sqrt{16 \pi^3 \omega_{\bm k}}} 
        ( u_s(\bm k) \hat{b}_{s}(\bm k) e^{-\ii k \cdot x} 
        \nonumber \\
        &+&
        v_s(\bm k) \hat{c}_{s}^\dagger(\bm k)  e^{\ii k \cdot x}) ,
        \label{expansion}
    \end{eqnarray}
where $u_s(\bm k)$ and $v_s(\bm k)$ are positive and negative frequency solutions with helicity $s$ and momentum $\bm{k}$  of the Dirac equation (see, {\em e.g.}, Ref.~\cite{Vanzella2}) and $\hat{b}_{s}(\bm k)$ and $\hat{c}_{s}(\bm k)^\dagger$ are creation and annihilation operators of particles and antiparticles, respectively, satisfying
$$
\{
\hat{b}_s (\boldsymbol{k}),\hat{b}^\dagger_{s'} (\boldsymbol{k}')
\} 
=
\{
\hat{c}_s (\boldsymbol{k}),\hat{c}^\dagger_{s'} (\boldsymbol{k}')
\} 
=
\delta^3 (\boldsymbol{k} -\boldsymbol{k}') \delta_{s,s'}.
$$
Using the field expansion from Eq.~(\ref{expansion}), we obtain
\begin{align}\label{neutronprotonexpansion}
    \hat{\bar{n}}\gamma_\mu (1 - \gamma^5)\hat{p} 
    &= \dfrac{1}{(2\pi)^3}\sum_{s, s'}\int \dfrac{\dd^3 k}{\sqrt{ 2\omega_{n}(\boldsymbol{k})}}\dfrac{\dd^3 k'}{\sqrt{2\omega_{p}(\boldsymbol{k'})}}
    \nonumber\\
    &\times e^{\ii \left[(\omega_{n}(\boldsymbol{k}) - \omega_{p}(\boldsymbol{k'}) )t - (\boldsymbol{k} - \boldsymbol{k'})\cdot\boldsymbol{x}\right]} 
    \nonumber\\ 
    &\times \left[\bar{u}_{s, n}(\boldsymbol{k})\gamma_\mu (1 - \gamma^5)u_{s', p}(\boldsymbol{k'})\right]
    \nonumber\\
    &\times \hat{b}^\dagger_{s, n}(\boldsymbol{k})\hat{b}_{s', p}(\boldsymbol{k'}) + \dots,
\end{align}
where we only display the term that contains the creation of a neutron and annihilation of a proton, corresponding to the nucleon excitation.

One can heuristically think of the sum over momenta in Eq.~\eqref{neutronprotonexpansion} as implementing a spatial profile for the proton-neutron system, which is assumed to be localized at the atom's location. Thus, the nucleon will be effectively pictured  as a  nonrelativistic quantum system following a well-localized spatial trajectory. This can be modeled by a classical current $j^\mu(x)$ with support on the nucleon with its quantum nature being encompassed by the internal degree of freedom of a two-level system. As a result, we replace the positively-charged hadronic current in Eq.~(\ref{FermiInt}) as follows:
\begin{equation}\label{proposal}
    \hat{\bar{n}}\gamma_\mu (1 - \gamma^5)\hat{p} \rightarrow j_{\mu}(x)e^{\ii \Delta M \tau}\hat{\sigma}^+,
\end{equation}
where $\Delta M\equiv M_n-M_p$ with $M_n$ and $M_p$ being the neutron and proton masses, respectively, $\tau$ represents the proper time associated with the center-of-mass trajectory, and $\hat{\sigma}^+$ is the two-level-system raising operator. Hence, the interaction Lagrangian~(\ref{FermiInt}) becomes 
\begin{align}\label{preliminarinteractionlagrangian}
\mathcal{L}_I 
&= -\dfrac{G_F}{\sqrt{2}}j_{\mu}(x)[e^{\ii \Delta M \tau}
\hat \sigma^+  \left(\hat{\bar{\nu}}_e\gamma^\mu (1 - \gamma^5)\hat{e}\right)
\nonumber\\
&+ e^{-\ii \Delta M \tau}
\hat \sigma^-\left(\hat{\bar{e}}(1 + \gamma^5)\gamma^\mu \hat{\nu}_e\right)].
\end{align}
The effect of taking the more complete effective Lagrangian for nuclear $\beta$-decay from~\cite{betadecay} here would simply amount to a refined expression for the current $j_\mu(x)$, with no conceptual impact in our discussion from now on. We note that we are not including any degree of freedom corresponding to the nucleon helicity and, thus, we do not consider here processes involving exchanges of angular momentum.

Now, we take a step further and expand the electron field using Eq.~(\ref{expansion})~\footnote{Equation~(\ref{expansion}) is written in the inertial frame which is at rest with the nucleon and, thus, $t \to \tau$.} to include it in the detector model. By doing so, we can write
\begin{align}\label{leptonic current} 
    e^{\ii  \Delta M \tau}(1 - \gamma^5)\hat{e} 
    = &\sum_{s = \pm 1}\int \dfrac{\dd^3 k}{\sqrt{(2\pi)^3 2\omega_{e}(\boldsymbol{k})}}
    \nonumber\\
    \times & e^{i\left[M_n - M_p - \omega_{e}(\boldsymbol{k})\right]\tau}
    e^{\ii \boldsymbol{k}\cdot\boldsymbol{x}}\hat{\psi}_{s,e}(\boldsymbol{k}) + \dots,
\end{align}
where we have defined
\begin{equation}
        \hat{\psi}_{s,e}(\boldsymbol{k}) \equiv (1 - \gamma^5)u_{s, e}(\boldsymbol{k})
        \hat{a}_{s, e}(\boldsymbol{k})
\end{equation}
and we only display the term associated with the nucleon ``excitation'':
\begin{equation}\label{nucleonexcitation}
 p + e^- + \bar{\nu}_e  \rightarrow n.
\end{equation}
It is important to note that, for each electron mode with momentum $\boldsymbol{k}$, the exponential time dependence in Eq.~(\ref{leptonic current}) strongly suggests we interpret $\Omega \equiv M_n - M_p - \omega_e(\boldsymbol{k})$ (including the energy carried by the electron) as being the effective energy gap for the fermion detector. In this sense, the presence of an auxiliary fermion at the detection (see Fig.~\ref{FeynmanFER}) allows us to see our fermion detector as a collection of UDW detectors with different energy gaps. 

By focusing on one particular energy gap, we use Eq.~(\ref{preliminarinteractionlagrangian}) to write our detector-neutrino interaction action as 
\begin{align}
    S_I 
    &= -\dfrac{G_F}{\sqrt{2}}\int \dd^4x\sqrt{-g} j_{\mu}(x)[\hat{\sigma}^+(\tau) \hat{\bar{\nu}}_\alpha\gamma^\mu \hat{\psi} 
    \nonumber\\
    &+ \hat{\sigma}^-(\tau)\hat{\bar{\psi}} \gamma^\mu\nu_\alpha],
    \label{semifinalmodel}
\end{align}
where $\hat{\sigma}^\pm(\tau) = e^{\pm \ii \Omega\tau}\hat{\sigma}^\pm$, as in the scalar case. We have replaced $e \to \alpha$, since the detector can be used to measure any neutrino. It is important to keep in mind that hereafter the energy gap $\Omega$  will include the lepton energy that participates in the process.  Here,
\begin{equation}\label{psizinho}
    \hat{\psi} \equiv \sum_{s=\pm 1}\!\!\!\!\underset{\:\:\:\:\:S(k_0, \delta k)}{\int}\dfrac{\dd^3 k}{\sqrt{(2\pi)^3 2\omega_{\alpha}(\bm{k})}}(1-\gamma^5) u_{s,\alpha}(\boldsymbol{k}) \hat{a}_{s,\alpha}
\end{equation}
where $S(k_0,\delta k)$ is a spherical shell centered at radius $k_0$ and with thickness $2\delta k$. $\hat{\psi}$ 
encodes the spinorial nature of the charged lepton which couples to the neutrino in the process. This is a left-chiral spinor operator, where  $u_{s,\alpha} (\boldsymbol{k})$ is the Dirac spinor computed at the central value of the lepton-$\alpha$ momentum distribution (with dispersion $\delta k$), and $\hat{a}_{s, \alpha}$ is the corresponding annihilation operator. (Note also that $\hat{\psi}$ is ``nondynamical" in the sense that the lepton dynamics has already been accounted for in the definition of the effective energy gap.)

A convenient orthonormal basis for the Hilbert space associated with our fermion detector  is
$$
\{ \ket{g}\ket{0}, \; \ket{g}\ket{1},\; \ket{e} \ket{0},\; \ket{e} \ket{1}\}, 
$$
where $\hat{\sigma}^\pm (\tau)$ acts on $\ket{g}$ and $\ket{e}$, as in the scalar case, and \footnote{ The $\Psi$ value will depend on the detector itself as well as different UDW detectors will have in general different $\Xi$ values:
$\hat{\sigma}^+ \ket{g} \equiv \Xi \ket{e},\;\hat{\sigma}^- \ket{e} \equiv \bar{\Xi} \ket{g}$. We have chosen $\Xi= 1$ to write Eq.~(\ref{sigma+-(tau)}) only because this the standard choice. We decided to proceed differently for $\Psi$ here to leave the physics of our fermion detector as transparent as possible. Moreover, as $\Xi$ in the scalar case, $\Psi$ will be eventually absorbed by the coupling constant, which shall be eventually measured.}
\begin{equation}\label{PSI}
\hat{\bar{\psi}} \ket{0} \equiv \bar{\Psi} \ket{1}, \; \hat{\psi} \ket{1} \equiv {\Psi} \ket{0}, 
\; \hat{\bar{\psi}} \ket{1} = \hat{\psi} \ket{0} \equiv 0. 
\end{equation}

Equation~(\ref{semifinalmodel}) comprises the interaction action of our detector with the fermionic field. For our purposes, it will be enough to consider the particular case where it is strictly pointlike. In this case,   
\begin{equation}
    j^\mu(x) = \dfrac{\delta^{(3)}(x - \mathsf{x}(\tau))}{\sqrt{-g}u^0}u^\mu(\tau),
\end{equation}
with $u^\mu(\tau)$ being the detector's 4-velocity. As a consequence, the interaction action~(\ref{semifinalmodel}) becomes
\begin{align}\label{finalmodel}
    S_I 
    &= -\lambda \int_{-\infty}^{+\infty }
    \dd\tau\, u_\mu(\tau) \chi (\tau)  
    \nonumber \\
    & \times  \left[\hat{\sigma}^+(\tau) \hat{\bar{\nu}}_\alpha (\mathsf{x}(\tau))\gamma^\mu \hat{\psi}  
    + \hat{\sigma}^-(\tau) \hat{\bar{\psi}}\gamma^\mu \hat{\nu}_\alpha (\mathsf{x} (\tau))\right],
\end{align}
where we have introduced the switching function $\chi(\tau )$, as in the scalar case, and $\lambda = G_F/\sqrt{2}$ (for a weakly-interacting detector). (Compare Eq.~(\ref{finalmodel}) with its scalar counterpart, Eq.~(\ref{SIE2}).)


\subsection{Fermionic oscillation probability}
\label{Fermionic oscillation probability}

We will now apply the results of Sec.~\ref{model}  to investigate the neutrino oscillation phenomenon with our fermion detector (which includes spinorial degrees of freedom absent in the simplified calculation of Sec.~\ref{scalarneutrinos}). We denote the trajectories of detector and source in the same way we did in Eq.~\eqref{inertialtrajectories}. The interaction action for a system with source, detector, and neutrinos is now given by 

    \begin{align}\label{fermionoscillationaction}
        &S_I =
        \nonumber \\
        & -\lambda_s \int \dd \tau_s u_\mu^{(s)}(\tau_s) \chi_s(\tau_s) \hat{\sigma}^-_s(\tau_s) \hat{\bar{\psi}}_s \gamma^\mu \hat{\nu}_{\alpha}(\mathsf{x}_s(\tau_s)) + \textrm{H.c.}
        \nonumber \\
        & -\lambda_d \int \dd \tau_d u_\mu^{(d)}(\tau_d) \chi_d(\tau_d) \hat{\sigma}^-_d(\tau_d) \hat{\bar{\psi}}_d \gamma^\mu\hat{\nu}_\beta(\color{black}\mf{x}_d\color{black}(\tau_d)) + \textrm{H.c.},
    \end{align}
which is the fermionic counterpart of Eq.~(\ref{interactionscalar2}), and  
    \begin{align}
        \hat{\nu}_\alpha(\mf x) = \sum_j U_{\alpha j} \hat{\nu}_j(\mf x).
    \end{align}
We will take the particular case where source and detector are inertial and at rest with respect to each other, as in~Eq.~\eqref{inertialtrajectories}. \textcolor{black}{In this case, the $4$-velocities of source and detector in the cartesian frame where both are stationary are given by}
    \begin{equation}
        u_\mu^{(s)}(t) = u_\mu^{(d)}(t) = (1, \bm{0}).
    \end{equation}

The interaction Hamiltonian comes from Eq.~\eqref{fermionoscillationaction}:
    \begin{align}\label{HsHd}
       \hat{H}_I(t) 
       & = \underbrace{ \lambda_s  \chi_s(t) \hat{\sigma}^-_s(t) \bar{\psi}_s \gamma^0  \hat{\nu}_\alpha(\mathsf{x}_s)+\textrm{H.c.}}_{\hat{H}_s(t) }\nonumber\\
        &\underbrace{+\lambda_d  \chi_d(t) \hat{\sigma}^-_d(t) \bar{\psi}_d \gamma^0 \hat{\nu}_\beta(\mathsf{x}_d)+\textrm{H.c.}}_{\hat{H}_d(t)}.
    \end{align}

\begin{figure}
\begin{center}
\vskip 1 truecm
\includegraphics[width=8 truecm]{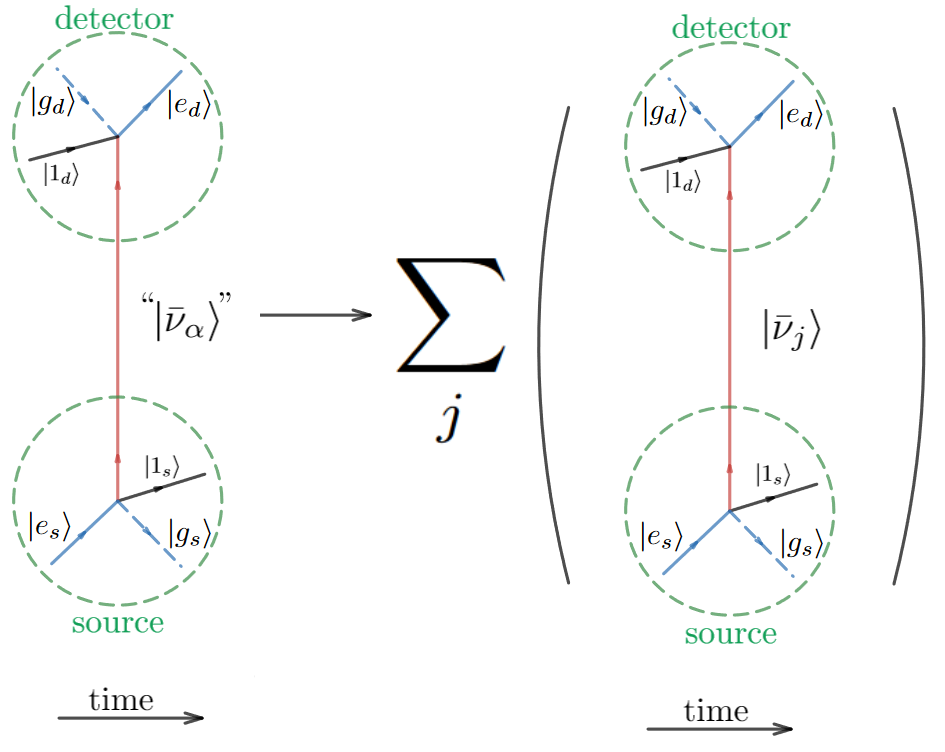}
\end{center}
\caption{The fermionic case is analogous to the scalar one except for the fact it includes extra fermionic lines associated with the leptons which turn out to be part of our fermion detector. The detector deexcitation associated with the neutrino emission at the source, 
$ |e_s\rangle \to (|g_s  \rangle + |1_s\rangle) + ``|\bar{\nu}_\alpha \rangle"$, 
should be compared with $ n \to (p + e^-) + \bar{\nu}_e$ and the aftermost neutrino detection, $ (|g_d \rangle + |1_d \rangle) + ``|\bar{\nu}_\alpha \rangle " \to |e_d \rangle $, should be compared with $ (p + e^-) + \bar{\nu}_e \to n $.
 } 
\label{FeynmanFER}
\end{figure}

The initial and final states associated with emission of a neutrino and aftermost detection is
    \begin{align}
        \ket{i} &= \ket{0} \Big(\ket{e_s}\ket{0_s}\Big)\Big(\ket{g_d}\ket{1_d}\Big),\\
        \ket{f} &= \ket{0} \Big(\ket{g_s}\ket{1_s}\Big)\Big(\ket{e_d}\ket{0_d}\Big),
    \end{align}
where $\ket{0}$ codifies that the neutrino only enters as a propagator line and $\ket{1_{s}}$, $\ket{1_{d}}$ and $\ket{0_{s}}$, $\ket{0_{d}}$ encode the presence and absence of the charged lepton coupling to the neutrino at the source and detector, respectively -- see Fig.~\ref{FeynmanFER}.

Now, we can compute the oscillation amplitude in complete analogy to the scalar case. Up to the lowest nontrivial order of perturbation theory, we have
    \begin{align}\label{fermionicintermetiateoscillationaux}
            \mathcal{A_{\alpha\rightarrow\beta}} 
             &=  - \bra{f}\mathcal{T}\exp(iS_I)\ket{i} 
            \nonumber \\
             &= \! - \!\! \int_{-\infty}^{+\infty}\!\!\!\!\!\! \dd t  \int_{-\infty}^{t}\!\!\!\!\!\! \dd t' \bra{f}\!(\hat{H}_s(t)\hat{H}_d(t') \! + \! \hat{H}_d(t)\hat{H}_s(t'))\!\ket{i}. 
    \end{align}
In order to evaluate it, we use Eq.~(\ref{HsHd}):
    \begin{align}\label{fermionicintermetiateoscillation}
            \mathcal{A_{\alpha\rightarrow\beta}}
            & = -\lambda_s\lambda_d \sum_{j}U_{\alpha j}U_{\beta j}^* 
            \int \dfrac{\dd^3p_j}{(2\pi)^3}\dfrac{1}{2\omega_j(\boldsymbol{p})}
            \nonumber \\
            & \times
            \Psi_s^\dagger\Big(
            F_j(\bm{p})(\slashed{p}_j+m_j) e^{-\ii \bm{p}\cdot\bm{L}} 
            \nonumber \\
            & - G_j(\bm{p})(\slashed{p}_j -m_j)e^{\ii \bm{p}\cdot\bm{L}}\Big)
            \gamma^0 \Psi_d,
    \end{align}
where  
$ \slashed{p}_j \equiv  \omega_j(\boldsymbol{p})\gamma^0 - \boldsymbol{p}_j \cdot\boldsymbol{\gamma}$, functions
$F_j(\bm{p})$ and~$G_j(\bm{p})$ are defined in Eqs.~\eqref{defF} and~\eqref{defG}, respectively, and 
    \begin{equation}
        \bar{\Psi}_s = \bra{1_s}\hat{\bar{\psi}}_s\ket{0_s},\;
        \Psi_d = \bra{0_d}\hat{{\psi}}_d\ket{1_d}
    \end{equation}
come from Eq.~(\ref{PSI}). It is worthwhile to note that in order to get Eq.~\eqref{fermionicintermetiateoscillation}, we have used
        $\bra{1_s}\psi_s\ket{0_s} = 0$ and $
        \bra{0_d}\bar{\psi}_d\ket{1_d} = 0$
(which makes several contributions in the Dyson expansion vanish) and the two-point functions
     \begin{align}
        \bra{0} \hat{\nu}_j(x)\hat{\bar{\nu}}_k(x')\ket{0}\! 
        &= 
        \! \frac{\delta_{jk}}{(2\pi)^{3}} 
        \!\! \int \!\! \frac{\dd^3 p_j}{{ 2\omega_j(\bm p)}} 
        (\slashed{p}_j\! + \! m_j) e^{\ii p_j \cdot (x-x')},\\
        \bra{0} \hat{\bar{\nu}}_k(x')\hat{\nu}_j(x)\ket{0}\! 
        &= 
        \!\frac{\delta_{jk}}{(2\pi)^{3}} 
        \!\! \int \!\! \frac{\dd^3 p_j}{{ 2\omega_j(\bm p)}} 
        (\slashed{p}_j\! -\! m_j) e^{-\ii p_j \cdot (x-x')}.
    \end{align}
We can now make further progress with Eq.~\eqref{fermionicintermetiateoscillation} by recalling that both $\Psi_d$ and $\Psi_s$ have left chirality~\footnote{Note that whenever the fermions $\Psi_1$ and $\Psi_2$ have the same chirality, the term proportional to the identity vanishes since $\Psi_1^\dagger\gamma^0\Psi_2 = \bar{\Psi}_1\Psi_2 = 0$.}:
    \begin{eqnarray}
        \mathcal{A}_{\alpha\rightarrow\beta}  
        &=& \lambda_s\lambda_d \Delta T \sum_j U_{\alpha j}U_{\beta j}^* 
        \int \frac{\dd^3 p_j}{(2\pi)^3}\frac{e^{\ii {\bm p}_j \cdot \bm L}}{\Omega_s^2 - \omega_j^2(\bm p)}
        \nonumber \\
        &\times&  \Psi_s^\dagger (\Omega_s \gamma^0 - {\bm p}_j \cdot \bm \gamma) \gamma^0 \Psi_d,\nonumber
    \end{eqnarray}    
where $F_j(\bm{p})$ and $G_j(\bm{p})$ were used as in Eqs.~\eqref{resultF} and~\eqref{resultG}, respectively.

Now, let us assume that either $\Psi_d$ or $\Psi_s$ have zero angular momentum in the direction of propagation. Using that our model assumes that all fermions are left-handed, it can be shown that the term $\bar{\Psi}_s({\bm p}_j \cdot \bm \gamma) \Psi_d$ does not contribute to the integral. Then,
    \begin{align}
        \mathcal{A}_{\alpha\rightarrow\beta} 
        & = 
        \lambda_s \lambda_d \Delta T \Omega_s 
        \Psi_s^\dagger \Psi_d 
        \sum_j U_{\alpha j}U_{\beta j}^* 
        \nonumber \\
        & \times \int \frac{\dd^3 p_j}{(2\pi)^3}
        \frac{e^{\ii {\bm p}_j \cdot \bm L}}{\Omega_s^2 - \omega_j^2(\bm p)}
        \nonumber \\
        & =  \ii \frac{\lambda_s\lambda_d}{4 \pi^2 L} \Delta T \Omega_s 
        \Psi_s^\dagger \Psi_d 
        \sum_j U_{\alpha j}U_{\beta j}^*  
        \nonumber \\
        & \times \int_{-\infty}^{+\infty} \dd p_j  \frac{ p_j e^{\ii p_j  L}}{\Omega_s^2 - \omega_j^2(p)}.
        \label{d3pfermion}
    \end{align}
    Equation \eqref{d3pfermion} is the fermionic analogue of Eq.~\eqref{d3pscalar}. By solving the integral as in the scalar case, we get
    \begin{equation}
        \mathcal{A}_{\alpha\rightarrow\beta} 
        = \frac{\lambda_{\text{eff}}^2}{4 \pi L}\Delta T \sum_j U_{\alpha j}U_{\beta j}^*  e^{\ii\Delta_j L},
    \end{equation}
where we have defined the effective coupling constant  $\lambda_{\text{eff}}^2 = \lambda_s\lambda_d\Omega_s \Psi_s^\dagger \Psi_d$. 
Notice that  $\lambda^2_\text{eff}$ in the fermionic case depends on the energy gap of the source in contrast to  $\lambda^2$ in the scalar case. This so because in the present fermionic case, $\lambda_s,\lambda_d \propto G_F $ have a dimension of energy$^{-2}$. By recalling \textcolor{pink}{,} that $\Psi_s$, $\Psi_d$ have dimension of energy$^{3/2}$ (see Eq.~(\ref{psizinho})) and $\Omega$ has dimension of  energy, we obtain that $\lambda_\text{eff}$ is dimensionless, as it should be.
    
It follows, then, that the probability of an ``$\alpha$-flavor'' neutrino to be detected as a ``$\beta$-flavor'' neutrino is
\begin{equation}\label{probabilidadealfabeta2}
        |\mathcal{A}_{\alpha\rightarrow\beta}|^2 
       \! =
       \! \dfrac{\lambda_\text{eff}^4}{16\pi^2 L^2 }
       \! \left[\dfrac{\sin{(\Delta\Omega\Delta t/2)}}{\Delta\Omega/2}\right]^2
       \!\left|\sum_{j}U_{\alpha j}U_{\beta j}^* e^{\ii \Delta_j L}\right|^2 \!\!\!,
\end{equation}
where we have used Eq.~\eqref{DeltaT}. The  corresponding detector excitation rate, for its turn, is obtained in the limit where the detector is turned on for an infinite  time interval, as in the scalar case: 
\begin{eqnarray}\label{finalprobfermion}
    \Gamma_{\alpha \to \beta} 
    &\equiv& 
    \lim_{\Delta t \rightarrow + \infty} 
    \dfrac{|\mathcal{A}_{\alpha\rightarrow\beta}|^2}{\Delta t}
    \nonumber \\ 
    &=& 
    \dfrac{\lambda_{\text{eff}}^4}{8\pi L^2}\delta(\Omega_d - \Omega_s)
    \left|\sum_{j}U_{\alpha j}U_{\beta j}^* e^{\ii \Delta_j L}\right|^2.
\end{eqnarray}
This is analogous to Eq.~\eqref{finalprobscalar} with the only difference being contained in the effective coupling constant. In contrast to the scalar case, $\lambda_\text{eff}$ will, in general, depend on the particular processes used at the source and detector to produce and observe the neutrinos. Of course, this information is washed out once  Eq.~(\ref{finalprobfermion}) is normalized as in Eq.~\eqref{finalescalar}. Indeed, the relative oscillation probability among different flavors is the same as in Eq.~(\ref{finalescalar}), except for the fact that in the fermionic case the PMNS matrix  elements can be complex.
    
\section{Conclusion}
\label{conclusion}
We have successfully described the phenomenon of neutrino oscillations without the need of flavor states, by using particle detectors to model the emission and absorption of neutrinos in charged-current weak interactions. Inspired by the UDW model, we first introduced most of the important conceptual elements and calculations in the framework of particle detectors by describing a simpler theory, where the flavor mixing occurred between real scalar fields. With this setup, we have shown how one can rephrase the standard neutrino oscillation probability formula in terms of processes directly associated with sources and detectors of neutrinos, with an area-law decay characteristic of isotropic emission by the source. Most importantly, the result is obtained without assuming any further knowledge about the propagating states for the fields. 
    
We then pursued a refinement of the UDW model, by including spinorial degrees of freedom that couple to neutrinos treated as fermionic quantum fields. We have shown how this can be motivated from physically reasonable simplifications of the $4$-fermion theory of weak interactions, which provides a good description of low-energy processes by which neutrinos are primarily emitted and absorbed. This demonstrates how we can naturally include important features of neutrinos that were left out in the scalar calculation, while remaining in the framework of localized particle detector models.
    
Under the assumption that the spin of the nucleons is unchanged in the processes of emission and absorption of neutrinos, the fermionic calculation can recover the same dependence on the distance and on the difference of squared masses of massive neutrinos that were derived in the scalar case and coincides with the usual quantum-mechanical calculation. This supports the intuition that particle detector models which couple two-level systems to scalar fields already capture much of the important physics in processes that do not involve exchanges of angular momentum between detector and field. 
 
Besides, the fermionic detector model devised in Sec.~\ref{model} also leaves new possibilities for future research. One such example would be the study of how the relative motion between source and detector may impact the oscillation phenomenon.
From the perspective of relativistic quantum information, the introduction of a new fermion-detector model also provides different venues for probing features of quantum fields that are not fully grasped by the case of scalar fields. 

Finally, our work vindicates Ref.~\cite{mixingneutrinos}, where the validity of the Unruh effect for mixing neutrinos was shown using solely mass states for the neutrinos. Therefore, in general, it is not true that one misses phenomena such as neutrino flavor oscillations by restricting oneself to a quantum description in terms of mass states (see, {\em e.g.}, Ref.~\cite{Blasone_2020,Petruzziello_2020} and reference therein). 
 
\section{Acknowledgements}
    
The authors are thankful to Andr\'e de Gouv\^ea, Marcelo Guzzo and Pedro Pasquini for helpful discussions. G.~E.~A.~M. is thankful to Gabriel Cozzella  for various conversations. Research at Perimeter Institute is supported in part by the Government of Canada through the Department of Innovation, Science and Economic Development Canada and by the Province of Ontario through the Ministry of Colleges and Universities. B.~S.~L.~T. and T.~R.~P. would like to acknowledge the Institute for Theoretical Physics of the S\~ao Paulo State University, South American Institute for Fundamental Research of the International Centre for Theoretical Physics (SAIFR-ICTP), Coordination for the Improvement of Higher Education Personnel (CAPES), and the University of Waterloo for financial support. G.~E.~A.~M. was partially supported by Conselho Nacional de Desenvolvimento Cient\'ifico e Tecnol\'ogico (CNPq) under Grant 301544/2018-2.
    
\bibliography{references}

\end{document}